\documentclass[final]{aipproc}

\layoutstyle{8x11double}

\newcommand{\be}{\begin{equation}}
\newcommand{\ee}{\end{equation}}
\newcommand{\ba}{\begin{eqnarray}}
\newcommand{\ea}{\end{eqnarray}}

\newcommand{\no}{\nonumber\\}

\def\tr{{\rm tr}}

\begin{document}

\title{Abnormal dilepton yield from parity breaking in dense nuclear matter}

\classification{25.75.-q, 25.75.Cj, 21.65.Jk, 12.40.Vv}
\keywords{Finite chemical potential, heavy ion collisions, local parity breaking}

\author{A.A. Andrianov, V.A. Andrianov}{
  address={V.A.Fock Department of Theoretical Physics,  Saint-Petersburg State University, 198504, 
St.Petersburg, Russia}
}

\author{D. Espriu}{
  address={CERN, Division PH-TH, 1211 Geneva 23, Switzerland.}
,altaddress={Departament d'Estructura i Constituents de la Mat\`eria and
Institut de Ci\`encies del Cosmos (ICCUB), Universitat de Barcelona, Mart\'\i ~i Franqu\`es 1, 08028 Barcelona, Spain.}
}

\author{X. Planells}{
  address={Departament d'Estructura i Constituents de la Mat\`eria and
Institut de Ci\`encies del Cosmos (ICCUB), Universitat de Barcelona, Mart\'\i ~i Franqu\`es 1, 08028 Barcelona, Spain.}
}

\begin{abstract}
At finite density parity can be spontaneously broken in strong interactions with far reaching implications.
In particular, a time-dependent pseudoscalar background would modify QED by 
adding a Chern-Simons term to the lagrangian.  
As a striking consequence we propose a novel explanation for the dilepton excess observed in heavy 
ion collisions at low invariant
masses. The presence of local parity breaking due to a time-dependent pseudoscalar condensate
substantially modifies the dispersion relation of photons and vector mesons propagating in such a medium,
changing the $\rho$ spectral function and resulting in a potentially large
excess of dileptons with respect to the predictions based in a `cocktail' of known
processes. 
\end{abstract}

\maketitle


\section{Introduction}

The appearance of parity violation via pseudoscalar condensation for sufficiently
large values of the chemical potential has been attracting much interest during 
the last decades to search it both in dense nuclear matter (in neutron/quark stars and
heavy ion collisions at intermediate energies) and in strongly
interacting quark-gluon matter (``quark-gluon plasma'' in heavy ion collisions
at very high energies). At finite baryon density was conjectured by A. Migdal in \cite{migdal}
long ago. While it was argued in \cite{witten} that
parity, and vector flavor symmetry could not undergo spontaneous symmetry breaking
in a vector like theory such as QCD, the conditions under which the results of
\cite{witten} hold (positivity of the measure) are not valid for non-zero
chemical potential.

We have investigated\cite{anesp} the possibility of parity being spontaneously violated in QCD at finite
baryon density and temperature. The analysis is done for an idealized homogeneous and infinite
nuclear matter where the influence of density can be examined with the help of constant chemical
potential. QCD is approximated by a generalized $\sigma$ model with two isomultiplets of scalars
and pseudoscalars. The interaction with the chemical potential is introduced via the coupling
to chiral quark fields as nucleons are not considered as point-like degrees of freedom
in our approach (for a semi-quantitative discussion this should suffice). This mechanism of
parity violation is based on interplay between lightest and heavy meson condensates
and it cannot be understood in simple models retaining the pion and nucleon sectors
solely; in particular is essentially different from the old idea of
pion condensation advocated originally by Migdal. We argue that, in the appropriate
environment (dense nuclear matter of a few normal densities where quark percolation does
not yet play a significant role), parity violation may be the rule rather than the exception.

Let us mention several  experimental signatures of parity violation 
in strong interactions\cite{anesp}:
a) Resonances do not have a definite parity and therefore the same resonance
can decay into even and odd number of pions.
b) At the very point of the phase transition leading to parity breaking one has
{\it six} massless pion-like states. After crossing the phase transition, in the parity broken
phase, apart from the usual pions, two scalar charged  states remain massless.
c) Changes in the
nuclear equation of state.
d) Additional isospin breaking effects in the pion decay constant and
substantial modification of $F_{\pi'}$.
However all these dramatic effects are nevertheless difficult to observe in the
environment of heavy ion collisions. We would like to find an effect that
showed up in simpler probes such as photons, electrons or muons.

During the last decade several experiments in heavy ion collisions  have indicated an abnormal 
yield of lepton pairs with
invariant mass $M < 1$ GeV in the region of small rapidities and moderate
transversal momenta \cite{ceres,phenix} (reviewed in \cite{lapidus,tserruya}).
This effect is visible only for collisions that are central or semi-central. 
From a comparison to $p p$ collisions it has been
established beyond doubt that such an
enhancement is a nuclear medium effect\cite{lapidus}.  
For the energies accessible at CERN SPS and
BNL RHIC the abnormal dilepton yield has not been yet explained
satisfactorily by known processes in hadronic physics\cite{lapidus,tserruya}.  
In this region 
the $\rho$ meson, directly via $\pi\pi$ fusion or indirectly
through the $\eta$ and $\omega$ Dalitz processes, largely dominates and the 
explanations based on in-medium effects of a dropping mass\cite{brown} and/or broadening resonance
seem unable to explain the spectacular dilepton enhancement\cite{heesrapp,renkr}.

We propose in this talk that the effect may be a manifestation of
local parity breaking (LPB) in colliding nuclei due to the generation of a pseudoscalar, isosinglet or
neutral isotriplet, condensate whose magnitude depends on the dynamics of the collision. It has been suggested
that such a background could be due to the topological charge fluctuations leading to the so-called
Chiral Magnetic Effect (CME)\cite{kharzeev} and seemingly
detected in the STAR and PHENIX experiments at RHIC\cite{star}, although
the issue is far from being settled. However the fact that the observed
dilepton excess is almost absent for peripheral collisions (where the CME should be more visible)
and maximized in central collisions makes us believe that it may be due to the ephemeral
formation of a bona-fide thermodynamic phase where parity
is broken\cite{anesp}.

\section{VDM and LPB}

It has been shown\cite{axion}  that an energetic photon propagating in 
this background may decay {\em on shell}
into dileptons. This same mechanism extended to vector mesons is proposed here as the source
for the abnormal dilepton yield in the range $ M < 1$ GeV,
for centrality $0 \div 20\%$ and for moderate $p_T < 1$ GeV. Here we will concentrate 
in the region around the $\rho$ and $\omega$ resonant contribution.

We shall assume that a time dependent but approximately spatially homogeneous background
of a pseudoscalar field $a(t)$ is induced at the
densities reached in heavy ion collisions and we will define a 4-vector related,
$\zeta_\mu  \simeq \partial_\mu a$. $a(t)$ could be either isosinglet or isotriplet
or even a mixture of the two, but detailed calculations will be presented for the case
of isosinglet background only.

The appropriate framework to describe
electromagnetic interactions of hadrons at low energies is the Vector Dominance Model (VDM)\cite{rapp,vmd}
containing the lightest vector mesons $\rho_0$ and $\omega$ in the $SU(2)$ flavor sector. 
We do not include $\phi$ meson,
as its typical mean free path $\sim 40$ fm makes it insensitive
to medium effects. Quark-meson interactions are described by
\be
\!\!{\cal L}_{int} = \bar q \gamma_\mu V^\mu q;\quad  V_\mu \equiv - e A_\mu Q  +
g_\omega  \omega_\mu \frac{\mathbf{I}}{2} + g_\rho \rho_\mu  \frac{\tau_3}{2},\label{veclagr}
\ee
where $Q= \frac{\tau_3}{2} + \frac16 \mathbf{I}$, $g_\omega \simeq  g_\rho \equiv g \simeq 6 $. 
The Maxwell and mass terms are
\ba
&&\!\!\!{\cal L}_{kin} = - \frac14 \left(F_{\mu\nu}F^{\mu\nu}+ \omega_{\mu\nu}\omega^{\mu\nu}+
 \rho_{\mu\nu}\rho^{\mu\nu}\right)\\ &&\!\!\!{\cal L}_{mass} =m^2_V \mbox{\rm tr}( V_{\mu}V^\mu) = \frac12  V_{\mu,a}  m^2_{ab} V^\mu_b, \no  &&\!\!\!\!\!\! m^2_{ab} =
m_V^2\left(\begin{array}{ccccc}
\frac{10 e^2}{9g^2} & &-\frac{e}{3g} && -\frac{e}{g} \\
 -\frac{e}{3g}&& 1 && 0 \\
 -\frac{e}{g} && 0 && 1 \\
\end{array}\right), \label{vdm}
\ea
with det $\left( m^2\right) = 0$ and 
$m_V^2 = m^2_\rho = 2 g^2_\rho f_\pi^2\simeq m^2_\omega$.  
Finally, in a pseudoscalar time-dependent background
the Lagrangian contains a parity-odd Chern-Simons (CS) term
\ba
&&{\cal L}_{CS}(k)\,= - \frac14 \varepsilon^{\,\mu\nu\rho\sigma}\, \tr{ \,\hat\zeta_\mu \, V_\nu(x)\, V_{\,\rho\sigma}(x)}\no
&&= \frac12 \tr{\,\hat\zeta \,\epsilon_{jkl}\, V_{j} \,\partial_k V_{l} }
= \frac12 \,\zeta \,\epsilon_{jkl}\, V_{j,a} \,N_{ab}\,\partial_k V_{l,b},
\ea
which additionally mixes photons and vector mesons due to LPB. For isosinglet pseudoscalar background
(the only possibility we shall consider here) $e^2 \hat\zeta=\frac95\zeta\mathbf{I}$, and
$N\propto m^2$.
Simple order-of-magnitude considerations indicate that $\zeta\sim \alpha
\tau^{-1} \sim 1 $ MeV, taking  the time of formation of pseudoscalar condensate $\tau= 1$ fm and the value of
condensate of order of $f_\pi$.

Particularizing to the
case $\zeta_\mu \simeq (\zeta, 0,0,0) $ we find the following spectrum
\ba
&&\!\!\!\! N\, = \, \mbox{\rm diag}\left[0,\,\frac{9g^2}{10 e^2},\, \frac{9g^2}{10 e^2}
+1\right] \sim \mbox{\rm diag} \left[0,\, 1,\, 1\right]\no
&& \!\!\!\! m^2\,  =\, m_V^2 \, \mbox{\rm diag} \left[0,\, 1,\,
1+ \frac{10 e^2}{9g^2}\right] ,
\ea
namely vector mesons have the dispersion relation
\be
k_0^2 - \vec k^2 = m_V^2 \pm \frac{9g^2}{10 e^2} \zeta |\vec k|\simeq m_V^2 \pm 360 \zeta |\vec k|\equiv m^2_{V,\pm} . \label{mvec}
\ee
Thus in the case of isosinglet background the massless photons are not distorted when
mixed with massive vector mesons.
In turn, massive vector mesons split into three polarizations with masses $m^2_{V,-} < m^2_{V,L}< m^2_{V,+}$ 
signifying local parity as well as Lorentz invariance breaking.
Note that the position of resonance poles for $\pm$ polarized mesons is moving with wave vector $|\vec k|$ and
therefore they reveal themselves as "giant" resonances. The enlargement of the resonant region potentially
leads to a substantial enhancement of their contribution to dilepton production away from their nominal
vacuum resonance position. See \cite{aaep} for more details.

\section{The NA60 results}

Here we shall limit our discussion to a comparison with the determination of the `abnormal' rho spectral 
function from the NA60 experiment\cite{ceres,na602,muonNA60}. The production rate of dileptons 
pairs mediated by $\rho$ mesons takes a form similar to the one
given in \cite{rapp} but
with modified propagators due to LPB, according to our previous discussion
\ba
\frac{dN}{d^4x dM}\simeq & \frac{c_\rho}{M^2}
\left (\frac{M^2-4 m_{\pi}^2}{m_{\rho}^2- 4 m_{\pi}^2}\right )^{3/2}
\sum_{\epsilon}\int_M^{\infty}dk_0\frac{\sqrt{k_0^2-M^2}}{e^{k_0/T}-1}\no
&\times \frac{m_{\rho,\epsilon}^4}{\left (M^2-m_{\rho,\epsilon}^2\right )^2+
m_{\rho,\epsilon}^4\frac{ \Gamma_\rho^2}{m_\rho^2}}.
\ea
Finite lepton mass corrections are not shown.
For $\omega$ mesons a similar expression is used but without the two pion threshold, characteristic of the
dominant coupling of the $\rho$ to hadronic matter. A simple thermal average is assumed
but we remark that the temperature $T$ is an effective one that may in fact depend on the 
range of $M$, $p_T$ and centrality.
The coefficients $c$ parameterize the total cross-sections for
vector meson creation. Because they are not known with precision in the present setting, particularly 
their off-shell values,
the relative weights are used as free parameters in the hadronic `cocktail'\cite{ceres,na602,muonNA60}. This is the
procedure used both by NA60 and PHENIX and we shall follow it here too. The usual `cocktail'
contains weights normalized to the peripheral collisions result (roughly agreeing with existing $pp$ and
$p$-A). For
semi-central and central collisions, particularly at low $p_T$ the $\rho/\omega$ ratio needs to be enhanced
by a factor 1.6 in the case of NA60 \cite{ceres} or approximately 1.8 in PHENIX \cite{rykov}.
\begin{figure}
  \includegraphics[scale=.16]{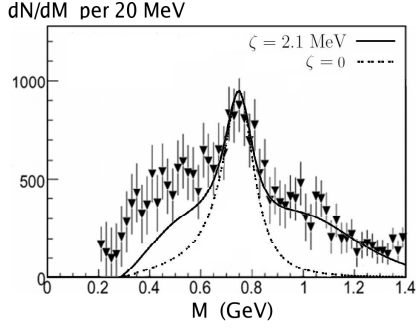}
  \caption{The $\rho$ meson contribution into the dilepton production is shown for parity symmetric nuclear
matter $\zeta = 0$ and for local parity breaking with $\zeta = 2$ MeV compared to the NA60 measurement
for central collisions and all $p_T$. Both
contributions are normalized to produce the same on-shell cross-section for $\rho$ meson production, which
is fitted to the experiment. The effective temperature is 300 MeV.}
\end{figure}

The result is shown in fig. 1 showing that LPB provides a much better description of the $\rho$ shape than 
the fitted hadronic cocktail. This can be further improved by including the (much smaller) modifications due
to LPB on the contribution from the $\omega$ pole and the $\omega$ Dalitz decay $\omega\to \mu^+\mu^-\pi^0$. 
This is tentatively shown in fig. 2. 

From these results we conclude that local parity breaking seems 
capable of explaining in a natural
way the PHENIX/CERES/NA60 `anomaly'. If LPB is confirmed (see \cite{aaep} for possible experimental signals)  
the consequences could be far reaching.
\begin{figure}
\includegraphics[scale=0.44]{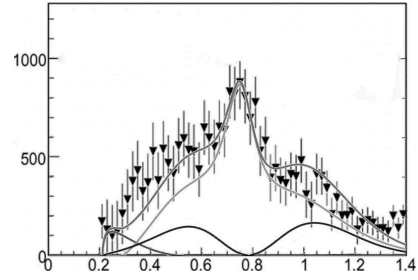}
  \caption{Same as in fig. 1 but including the smaller contributions from the LPB modifications to the $w$ pole and
Dalitz decay (shown separately), after adjusting the coefficients for an optimal fit. $\zeta=1.9$ MeV, $T=270$ MeV. The overall fit is 
excellent giving strong plausability to the possibility of LPB.}
\end{figure}



\bibliographystyle{aipproc}   

\end{document}